\documentclass[10pt,a4paper,oneside, titlepage]{article}\usepackage[]{graphicx}\usepackage[]{color}
\makeatletter
\def\maxwidth{ %
  \ifdim\Gin@nat@width>\linewidth
    \linewidth
  \else
    \Gin@nat@width
  \fi
}
\makeatother

\definecolor{fgcolor}{rgb}{0.345, 0.345, 0.345}

\usepackage{framed}
\makeatletter
 {\par\unskip\endMakeFramed%
 \at@end@of@kframe}
\makeatother

\definecolor{shadecolor}{rgb}{.97, .97, .97}
\definecolor{messagecolor}{rgb}{0, 0, 0}
\definecolor{warningcolor}{rgb}{1, 0, 1}
\definecolor{errorcolor}{rgb}{1, 0, 0}
\newenvironment{knitrout}{}{} 

\usepackage{alltt}

\usepackage[T1]{fontenc}
\usepackage[UKenglish]{babel}
\usepackage{natbib}
\usepackage[Export]{adjustbox}
\usepackage[parfill]{parskip}
\usepackage{amsmath}
\usepackage[pdfborder={0 0 0},
            colorlinks=true,
            urlcolor=blue,
            linkcolor=black,
            citecolor=black,
            filecolor=black]{hyperref}

\IfFileExists{upquote.sty}{\usepackage{upquote}}{}
\begin{document}

\title{The dynamically changing publication universe as a reference point in national impact evaluation: A counterfactual case study on the Chinese publication growth\footnote{This work was supported by German Federal Ministry of Education and Research (Grant 01PQ13001 and 01PQ17001)}}

\author{Stephan Stahlschmidt and Sybille Hinze\\
\\
German Centre for Higher Education Research\\
and Science Studies (DZHW)\\
\\
\texttt{\{stahlschmidt,hinze\}@dzhw.eu}}

\date{\vspace{0.25cm}\textit{WORKING PAPER}\\
\vspace{0.5cm}
\textit{Version: \today}}

\maketitle

\begin{abstract}
National bibliometric performance is commonly measured via relative impact indicators which appraise absolute national values through a global environment. Consequenty the resulting impact values mirror changes in the national performance as well as in its embedding. In order to assess the importance of the environment in this ratio, we analyse the increase in Chinese publications as an example for a structural change altering the whole database.

Via a counterfactual comparison we quantify how Chinese publications benefit a large set of countries on their impact values, identify explanatory factors and describe the underelying mechanism due to longer reference lists and a non-uniform citation distribution among recipient countries. We argue that such structural changes in the environment have to be taken into account for an unbiased measurement of national bibliometric performance.
\end{abstract}

\section*{China as a non-marginal influence on bibliometric indicators}

Since 2006 China counts as the second biggest single "producer" of publications trailing only the United States and consequently increasingly competes for the space in \textit{Web of Science} indexed journals. This growth and subsequent influence of Chinese publications is too extensive to be considered marginal, but affects the whole database. Furthermore its is stimulated by economic growth and political factors \citep{Zhou2016} rendering it an external influence on the science system. Consequently the question arises, how this unprecedented growth of contributions from a single country with its specific bibliometric characteristics affects the whole bibliometric measurement process.

The leverage of Chinese publications on non-Chinese ones is facilitated by the relative nature of the commonly applied impact indicators. The mean normalized citation score \citep{Waltman2011} or the share of highly cited papers \citep{Waltman2013} are relative indicators relating absolute values of a particular country to the whole publication universe. Consequently they mirror changes on both sides, the national one expressed in the numerator, as well as changes in the global  environment sumed up by the denominator. However common understanding relates any observed changes in the impact value to the country, the numerator, and not to the environment, the denominator.

We argue that the changes to the global environment need to be taken into account in the discussion more extensively and thus analyse the Chinese publication increase as an example of the dynamics governing the evolution of the database. Effects on national impact indicators and the underlying mechanism are empirically analysed and related back to the Chinese publication increase. I.e. it is shown how the denominator influences the observed values.

While former work on the rise of Chinese publications is often concerned with describing the impact of the Chinese appearance at the forefront of scientific publications, e.g. country shares of publications and citations \citep{Cote2016, Bornmann2013, Leydesdorff2009} or relating these bibliometric measures with socio-economic data \citep{May1997}, we are more interested in the arising measurement issues implied by the growth of Chinese publications. Consequently we will contrast the just described state of the universe of scientific publications with a counterfactual bibliometric universe without China. We analyse how developed countries would have performed under supposedly stable conditions and compare this counterfactual outcome with the actual one to deduce the effect Chinese publications exhibit on other countries.

This research comprises several implications for policy setting agents. We like to answer, to what extent the observed increase in bibliometric impact measures in the developed world might ultimately be assigned to country specific science programmes and policies and to what extent the improvement denotes an artificial measurement artefact caused by a dynamic environment and respectively the way the indicators are constructed which are used to reflect a country's scientific performance.

\begin{knitrout}
\definecolor{shadecolor}{rgb}{0.969, 0.969, 0.969}\color{fgcolor}\begin{figure}[t]

{\centering \includegraphics[width=6.5in,height=4.5in,center]{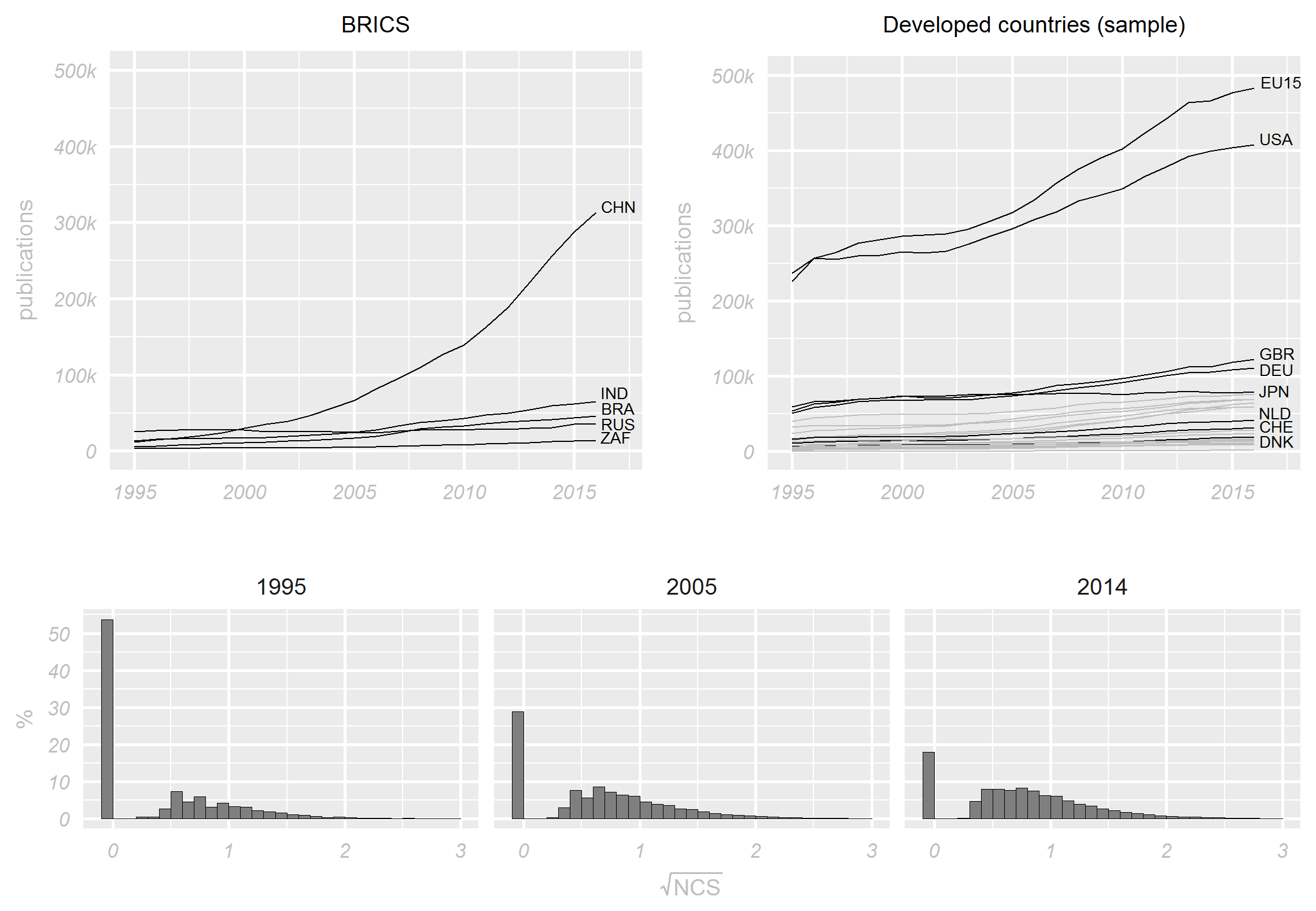} 

}

\caption[Growth of national publications among BRICS countries (upper left panel), national publication growth of a sample of OECD countries (upper right panel) and yearly histograms (lower panel) of the normalised citation score for Chinese publications (Data is truncated in the upper tail)]{Growth of national publications among BRICS countries (upper left panel), national publication growth of a sample of OECD countries (upper right panel) and yearly histograms (lower panel) of the normalised citation score for Chinese publications (Data is truncated in the upper tail).}\label{fig:CHN_bib_Graph}
\end{figure}

\end{knitrout}

The upper panels in Figure \ref{fig:CHN_bib_Graph}\footnote{Interactive versions of several graphs can be found at \href{http://bit.ly/2r8fV6j}{http://bit.ly/2r8fV6j}} illustrate the growth of national scientific contributions in the \textit{Web of Science} over the last 20 years. While the United States of America and the combined EU15 countries clearly dominate in this measure of productivity, China has observed a tremendous growth in its publication outlet. While still going along with its BRICS companions in 1995, it surpassed them in the year 2000 and all other countries apart from the United States of America six years later in 2006. Since its growth has not slowed down, but constantly exceeds the national counterparts of all other BRICS and OECD countries.

However the observed exponential growth of Chinese publications might affect bibliometric impact measures only if its citation characteristics differs from the worldwide citation distribution. If not, the Chinese publications only add further publications to the corpus of scientific publications allowing for more precision in the measurement process.

Figure \ref{fig:CHN_bib_Graph} depicts in the lower panel the citation distribution of Chinese publications, i.e. the year and subject field specific normalised citation score (NCS) of every Chinese article or review appearing in a \textit{Web of Science} indexed journal based on a triannual citation window. According to this illustration China has come a long way starting with median normalised citation score of 0 in 1995 to surpassing 0.5 in 2014. At the same time the $10\%$ best Chinese publications achieved a normalised citation score of at least 1.5 in 1995, respectively 2.4 in 2014.

But Chinese publications do not only differ on the cited side from global averages, but also on the citing side. Chinese publications include longer reference lists which subsequently result in changes to the globally applied discipline specific citation statistics. Furthermore its citations focus especially on Asian and scientifically advanced countries, granting them higher impact values. Consequently we argue that due to their size and bibliometric characteristics Chinese publications wield a non-marginal influence on the publication universe and alter the reference points of any national impact evaluation. Hence the observed changes in national impact values over time might not only describe the nationl performance alone, but also structural changes in the environment applied to measure the national performance.

The next section will explain the applied methodology of constructing a counterfactual bibliometric world without Chinese publications. Afterwards we will contrast this counterfactual setting with the actual one to observe and explain the effects Chinese publications exhibit. In the subsequent discussion we present a notional outlook and comment on the wider implications. The last section concludes.

\section*{A counterfactual bibliometric world without China}

Our analysis is based on the \textit{Web of Science} raw data\footnote{Data is provided by the German Competence Centre for Bibiliometrics (\href{www.bibliometrie.info}{www.bibliometrie.info}).} which details the citation links between all included scientific contributions. We limit our analysis of cited items to \textit{articles} and \textit{reviews} from the \textit{Web of Science core collection} sections \textit{Science Citation Index Expanded}, \textit{Social Sciences Citation Index} and \textit{Arts $\&$ Humanities Index}, as these journal publications constitute the primary communication device for new findings in most disciplines.

The coverage of the citing side, presenting the reception of the aforementioned \textit{articles} and \textit{reviews}, poses no such restriction on the document type and includes furthermore also the \textit{Conference Proceedings Citation Index -- Science} and \textit{Conference Proceedings Citation Index -- Social Sciences} to allow for an ample reception via the corresponding citations. Self-citations are understood to largely represent an essential part of scientific progress and are consequently included in the analysis. Citations to or from non-source items out of the scope of the \textit{Web of Science core collection} are omitted for obvious reasons.

In order to cover the apparent start of the rise of Chinese publications in the late 1990s we commence our analysis in 1995 and extend it to the most recent data. Allowing for a triannual citation window we report on impact indicators for publications up to 2014 while incorporating citations from up to 2016. These citations are normalized on the corresponding year and discipline of the cited publication. In doing so we employ the 252 \textit{Web of Science Subject Categories} and the publication year as stated in the database.

Apart from citation links the \textit{Web of Science} also includes affiliation data of the respective authors. The listed residence of the affiliation on a country level facilitates an identification of Chinese publications in the \textit{Web of Science} publication universe and therefore our study. While authorship of scientific papers, and especially the individual contributions of the listed authors to a paper, are widely discussed in the literature \citep{Biagioli2003}, we build our analysis upon the classical assumption, that any listed author has contributed an essential part to the paper. Consequently without that particular contribution the corresponding paper would not have been published. Applying this reasoning to the definition of Chinese publications, we assign any publication to the set of Chinese publications if at least one listed author is affiliated with an institution residing in China. While the contributions of all other co-authors without a Chinese affiliation is thererby striped of their institutional origins, the potential meassurement error introduced by this approach might substantially be lowered by noting that many of these co-authors without a Chinese affiliation are actually of Chinese lineage \citep{Wang2013} and consequently also embody the rising influence of China throughout the global science system.

Based upon this definition of Chinese papers we quantify the effect of this ever increasing stock of Chinese publications on the bibliometric universe by inferring what would have happened without these Chinese publications. Consequently a counterfactual bibliometric world without Chinese publications is constructed and applied as a placeholder for a stable bibliometric environment unaffected by the rise of Chinese publications. Contrasting this counterfactual with the actual bibliometric universe allows for an assessment of the effect Chinese publications exhibit on the bibliometric universe.

This approach borrows from the treatment effect literature in Economics \citep{Rubin1974, Imbens2009}, which itself builds upon John Stuart Mill's method of differences \citep{Mill1843}. Based upon observational data treatment effects models infer if and how a treatment, an exogenous stimulus, causally affects a target audience. Ideally these models compare the same observational units with and without the stimulus on some outcome variable and declare any difference in the output to denote a causal effect of the treatment. Obviously any unit can either be exposed or not be exposed to the treatment and a direct comparison on the same unit is infeasible \citep{Holland1986}. Consequently treatment effects models apply carefully constructed substitute comparisons exploiting the untreated units of the population. However, as the Chinese publications affect the whole \textit{Web of Science} publications universe and every country enlisted, no unaffected units are available, but have to be constructed artificially.

In doing so, we exclude the aforementioned set of Chinese publications from the cited and citing side of the \textit{Web of Science} publication universe and recount the citations from source items to the remaining non-Chinese publications. Afterwards these counts are applied to recompute the \textit{Web of Science Subject Categories} based expected counts \citep{Waltman2011} and $90\%$ quantile thresholds \citep{Waltman2013}. In the resulting counterfactual bibliometric setting we subsequently compare each non-Chinese publication to these counterfactual statistics to obtain national impact indicators.

Finally these averaged values are contrasted with the actual national values for country $i$ via
\begin{eqnarray}
\Delta Impact^{(i)} = Impact_{actual}^{(i)} - Impact_{counterfactual}^{(i)},\label{eq1}
\end{eqnarray}
where $Impact$ denotes either the national \textit{Mean Normalized Citation Score} (MNCS) or the national share of the $10\%$ most cited publications (PP(top10)). As the two bibliometric worlds differ only in the Chinese publications $\Delta Impact^{(i)}$ quantifies how national bibliometric impact indicators are affected by Chinese publications.

This comparison between the actual and counterfactual impact indicators is facilitated by the particular nature of the indicators, as they evaluate publications by comparing the respective citation counts with citations in a pre-determined environment. Consequently the resulting absolute decline in publications and citations imposed by the exclusion of Chinese publication might potentially affect both the publication specific count of obtained citations and the discipline specific citation statistics like expected citations or the PP(top10) thresholds. However, these level shifts do not inhibit the aforementioned comparison if we assume that the actual and counterfactual worlds truthfully describe the complete national publication output in the respective settings. Consequently any difference between the actual and counterfactual national impact indicators results soly from the change in the environment driven by Chinese publications.

This reasoning highlights a fundamental condition of our approach, namely such counterfactual analysis seems only feasible on a macro level. The increasing share of Chinese publications in the publication universe and the therein stated set of knowledge renders the identification of counterfactual actions by individual scientists infeasible and consequently the counterfactual state of single publications cannot be deduced. However, the not unusual case of parallel discoveries of important findings probably result in a comparable albeit potentially smaller knowledge set in the counterfactual world. Furthermore national contributions to this knowledge set seem to be underpinned by robust steady trends. Hence, we assume that national impact indicators in the counterfactual world do depict a realistic picture on a macro scale. Still the constructed counterfactual bibliometric world does not constitute a perfectly known alternative setting but rather expresses a \textit{Gedankenexperiement} -- a thought experiment -- which helps to understand the mechanism driving the evolution of the database and the resulting consequences for national impact indicators.

\section*{Comparing the counterfactual with the actual bibliometric world}

Having processed the counterfactual bibliometric world we may apply equation (\ref{eq1}) to observe the effect on a national level. Afore we may note the evolution of national MNCS for a set of OECD countries\footnote{Countries shown in graph: Australia, Austria, Belgium, Canada, Denmark, Finland, France, Germany, Great Britain, Greece, Ireland, Israel, Italy, Japan, Luxemburg, Netherlands, Norway, New Zealand, Portugal, South Korea, Spain, Sweden, Switzerland, United States of America} in the upper left panel in Figure \ref{fig:MNSCGraph}. In general the MNCS is steadily improving for nearly all listed countries with the notable exceptions of the United States of America and Japan. Three small European countries lead and are trailed by Great Britain as the first of a group of larger countries including Germany and France. The included Asian countries perform less well on this measure, while the big Oceanian countries line up with their European counterparts.

\begin{knitrout}
\definecolor{shadecolor}{rgb}{0.969, 0.969, 0.969}\color{fgcolor}\begin{figure}[t!]

{\centering \includegraphics[width=6.5in,height=4.5in,center]{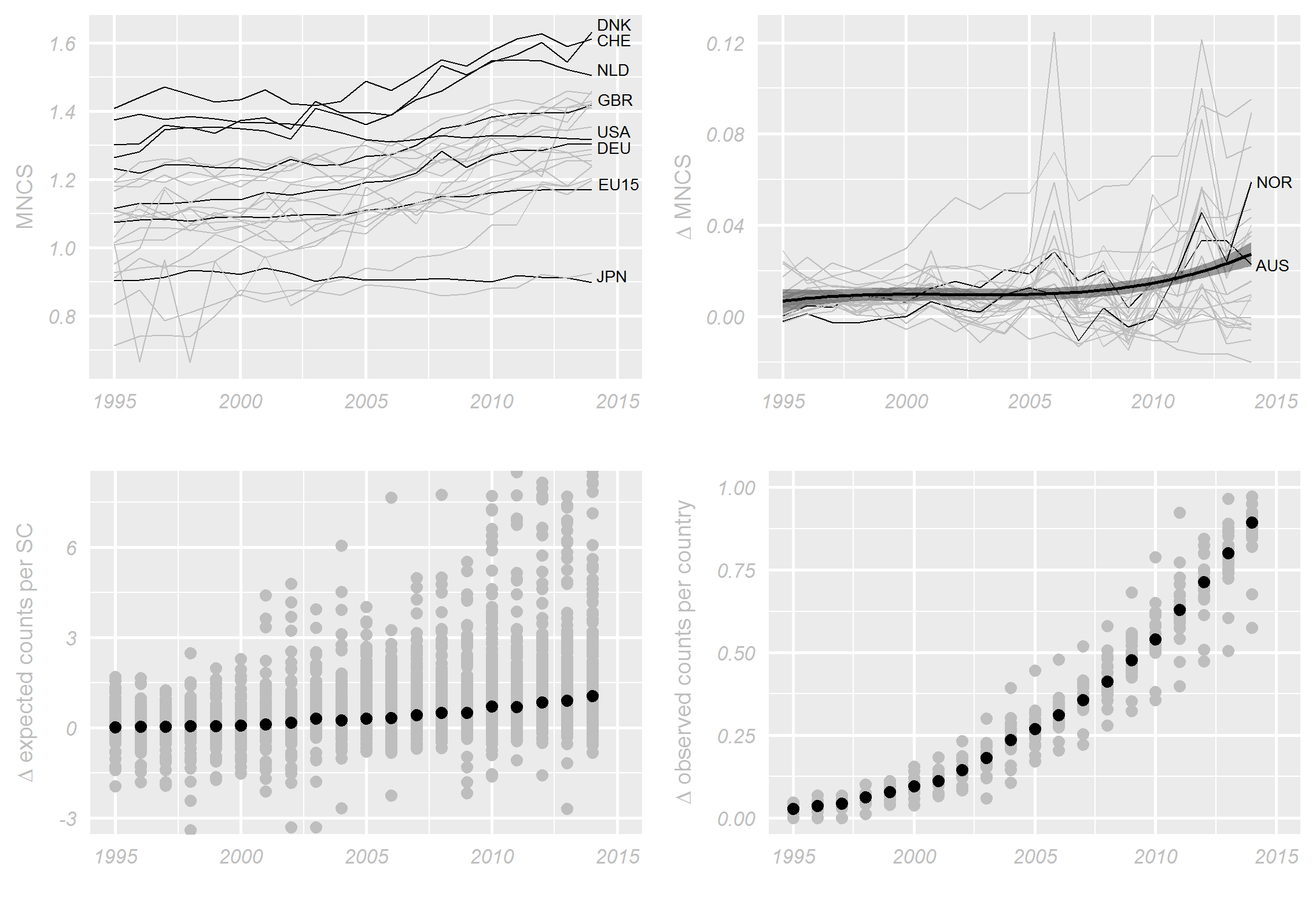} 

}

\caption[Actual national MNCS values (upper left panel), changes in MNCS (upper left panel) and differences in in expected counts (lower left panel), respectively observed citations (lower right panel) due to the increase in Chinese publications]{Actual national MNCS values (upper left panel), changes in MNCS (upper left panel) and differences in in expected counts (lower left panel), respectively observed citations (lower right panel) due to the increase in Chinese publications}\label{fig:MNSCGraph}
\end{figure}

\end{knitrout}

The top right panel in Figure \ref{fig:MNSCGraph} shows the effect China inhibits on these national MNCS as it illustrates the resulting differences $\Delta MNCS^{(i)}$ of equation (\ref{eq1}). Every grey line represents a single country $i$, while the black line represents an average of these national values accompanied by a confidence interval. As most values and especially the average values are positive we conclude that the MNCS values of the listed countries benefit from the Chinese publications. I.e. independent of their own performance these countries obtain larger MNCS values due to the inclusion of Chinese publications in the publication universe. This effect, albeit rather small in the beginning, increases over time and affects the single countries to a varying degree. Indeed the yearly variance of $\Delta MNCS^{(i)}$ over $i$ increases as time proceeds.

While the observed existence of this effect is enlightening in itself, we are especially interested in understanding how this effect accrues in the first place. However the model alone does not reveal, how the final effects arise, due to its rather flexible approach with a minimum of imposed parametric modelling structure. The comparison of the actual and counterfactual bibliometric world mirrors a black box method and the underlying mechanism has to observed via suitable descriptive statistics.

Therefor we compare the expected and obtained ciations between the actual and counterfactual setting as these values drive the MNCS of country $i$ with publications $j\in\left\{1,\ldots,J\right\}$ via
\begin{eqnarray}
MNCS^{(i)} = \frac{1}{J}\sum^{J}_{j=1} \frac{obtained~citations_j}{expected~citations_j}\label{eq2}
\end{eqnarray}
for each year. An increase in obtained citations and an decrease in expected citation will independently result in a rise of the national MNCS. Obviously Chinese publications might affect both citations counts, as these Sinic publications might cite publications of other countries whereby increasing the citation count of non-Chinese publications. Furthermore any citation in the closed world of the \textit{Web of Science} publication universe, and independently of its national origin, affects the overall citation count expressed in the yearly, \textit{Subject Categories}-specific expected citation counts.

The lower left panel of Figure \ref{fig:MNSCGraph} compares the expected counts $EC$ between the actual and counterfactual bibliometric world. Every grey dot denotes a difference $\Delta EC^{(h)}$ in the $h\in\left\{1,\ldots,252\right\}$ \textit{Subject Categories}:
\[
\Delta EC^{(h)} = EC^{(h)}_{actual} - EC^{(h)}_{counterfacutal}.
\]
The black dot marks the mean difference among all \textit{Subject Categories}. While in the beginning of the observation period no effect seems recognizable, this changes in later years as the $\Delta EC$ gradually increase in size and finally nearly all differences are positive. Hence the exclusion of Chinese publications lowers the overall expected counts or, the other way round, the inclusion of Chinese publications increases the overall standard of expected citations.

Having observed an increase in national MNCS and expected citation counts due to the Chinese publications, equation (\ref{eq2}) postulates also an increase in obtained citations. Indeed this increase in obtained citations must surpass the increase in expected counts for equation (\ref{eq2}) to hold. Accordingly the lower right panel in Figure \ref{fig:MNSCGraph} presents the average publication-level differences in obtained citations $OC$ for every publication $j$ of country $i$:
\[
\Delta OC^{(i)} = \frac{1}{J}\sum^{J}_{j=1}OC^{(j)}_{actual} - OC^{(j)}_{counterfacutal}.
\]
While the additional obtained citations in the beginning seem negligible $\Delta OC^{(i)}$ clearly increases over time for all countries. As before the variance over countries seems to increase as some countries benefit more than other from the additional citations brought in by Chinese publications.

While this observation of increasing obtained citations validates equation (\ref{eq2}), these changes only represent empirical symptoms of the underlying mechanisms. In order to analyse these underlying implications of the increase in Chinese publications we start with a highly stylized bibliometric toy model. In detail we omit for a moment any citations across time or disciplines and any influence non-source items might have. In such a perfectly encapsualted setting citations are distributed as a zero-sum game from reference lists and the expected citation count equals the average reference list length, as every reference refers to a particular source item in the database. Consequently any expansion in terms of additional publications might only increase (decrease) the expected citation counts if the these additional publications include more (less) than usual references. Furthermore if citations were to be split up in fractions, a country with a longer than usual reference lists could enter neutrally in the database without distorting it. Therefor every country would need to be compensated the exact amount of citations necessary to keep the same MNCS value, although the expected citation counts rise due to the longer reference list. The remaining citations from the additional reference lists would be allocated to the newly entering country itself and would perfectly match the new expected counts leaving that country with a MNCS value of 1. This mechanism mirrors the (neo-)liberal economic theory, that printing money does not affect the economy, but only leads to uniform price increases.

In order to illustrate this point we depict the counterfactual reference list lengths and counterfactual expected citation counts for all \textit{Web of Science Subject Categoris} in the year 2014 in the upper left panel of Figure \ref{fig:expectedcounts_Graph}. As indicated by the toy model a profound positive relation might be indentified in which longer reference lists are accompanied by higher expected citaion counts. However this influence of the reference list length on the expected citation count is obscured by the frequent citation links across time and the less frequent citation links across disciplines. Furthermore the aforementioned enlarged base of citing publications interferes in the direct relation between reference list lengths and expected citation counts. Summing up Chinese publications may only affect expected citation counts if they differ in reference list length, citation links over time, citation links across disciplines or non-source items.

\begin{knitrout}
\definecolor{shadecolor}{rgb}{0.969, 0.969, 0.969}\color{fgcolor}\begin{figure}[t!]

{\centering \includegraphics[width=6.5in,height=4.5in,center]{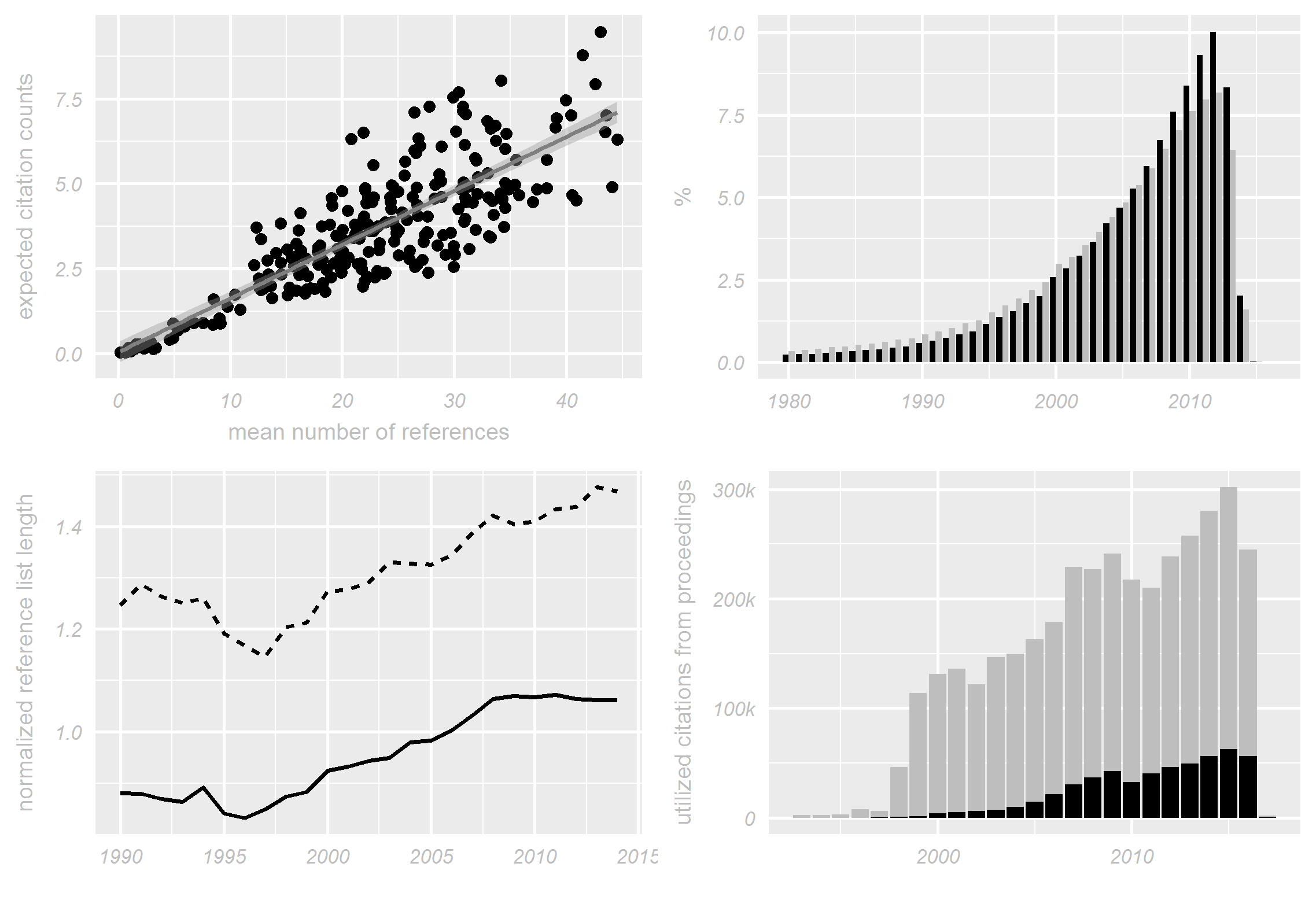} 

}

\caption[Scatterplot of Subject Categories concerning their average reference list length and expected citation counts in 2014 (upper left panel), normalized length of Chinese reference lists (solid line) and normalized length of Chinese reference lists (dashed line) utilized in citation window (lower left panel), distribution of cited article over time from citing articles from 2014 (upper right panel) and stacked yearly count of utilized citations from non-standard document types (lower right panel)]{Scatterplot of Subject Categories concerning their average reference list length and expected citation counts in 2014 (upper left panel), normalized length of Chinese reference lists (solid line) and normalized length of Chinese reference lists (dashed line) utilized in citation window (lower left panel), distribution of cited article over time from citing articles from 2014 (upper right panel) and stacked yearly count of utilized citations from non-standard document types (lower right panel). Besides upper left panel Chinese values are depicted in black, while grey values denote the sample-based, non-Chinese equivalents.}\label{fig:expectedcounts_Graph}
\end{figure}

\end{knitrout}

The solid lines in the lower left panel of Figure \ref{fig:expectedcounts_Graph} depicts the average of normalized reference list lengths of Chinese publications for the last 25 years. We normalize Chinese reference lists on their worldwide non-Chinese counterparts as countries differ in their disciplinary focus \citep{EC1997} while concurrently disciplines can be distiniguished by their reference behaviour foiling a comparison on absolute values. According to the graph Chinese reference lists have been shorter than the global average in the beginning of the observation period. A substantial growth started in the end of the 1990s and apparently stopped ten years later. Still during this period the average normalized Chinese reference list has exceeded the world average ever since 2006. Hence the observed increase in expected citation counts is at least since 2006 influenced by the Chinese reference lists.

However the observed increase in expected citation counts begins much earlier and must therefore partly be explained by the other interfering factors. As an increase in expected citation counts can be observed uniformly across all \textit{Subject Categories}, Chinese publication seem unlikely to differ strongly in their use of citations across disciplines. Contrary citations crossing time periods are common and the varying time focus of Chinese and non-Chinese publications can exemplarily be observed for citing publications from 2014 in the top right panel of Figure \ref{fig:expectedcounts_Graph}. Chinese publications focus much stronger on more recent publications, as their share of references to publications from 2007 to 2014 surpasses the non-Chinese shares and trails them for all preceding years.

At the same time the applied triannual citation window curtails the count of relevant citations to cited publications not older than two years than the citing publication and consequently favours the Chinese focus on more recent literature. Accordingly the dashed lines in the lower left panel in Figure \ref{fig:expectedcounts_Graph} depicts the same comparison of reference list length as before but restricts the count to cited publication within the triannual citation window. Comparing the two lines it can be observed that Chinese reference lists exceed the non-Chinese ones in terms of citations utilized via the citation window for the whole observation period. Most recently Chinese publications include nearly 50$\%$ more utilized citations and raise the global expected citation counts accordingly.

A further, more technical explanation of the increase in expected citation counts stems from our particular definition of the citing side. While our analysis restricts the cited side to \textit{articles} and \textit{reviews} the citing side includes also other document types including \textit{proceedings papers}. The lower right panel in Figure \ref{fig:expectedcounts_Graph} depicts the number of additional Chinese citations utilized in our citation counts stemming from this extended defintion of the citing side. While comparably small in size a substantial growth in these citations can be identified. Hence the expected citation counts rise as the set of citations from these non-standard document types also includes a growing share of Chinese items.

In general any country will benefit from Chinese publications if the additional citations received will outweigh the rise in the expected counts driven by the longer reference lists. Two factors might influence to what extend other countries will receive additional citations from Chinese publications.  First an outward looking China, which cites foreign papers relatively more often than national publications. In this case Chinese publications do not only have to keep up with the expected counts set by other countries but also suffers from the self-imposed increase in those standards. Second a non-uniform spread of citations among countries, where disproportionally many citations are received by a particular group of countries.

\begin{knitrout}
\definecolor{shadecolor}{rgb}{0.969, 0.969, 0.969}\color{fgcolor}\begin{figure}[t!]

{\centering \includegraphics[width=4in,height=4in,center]{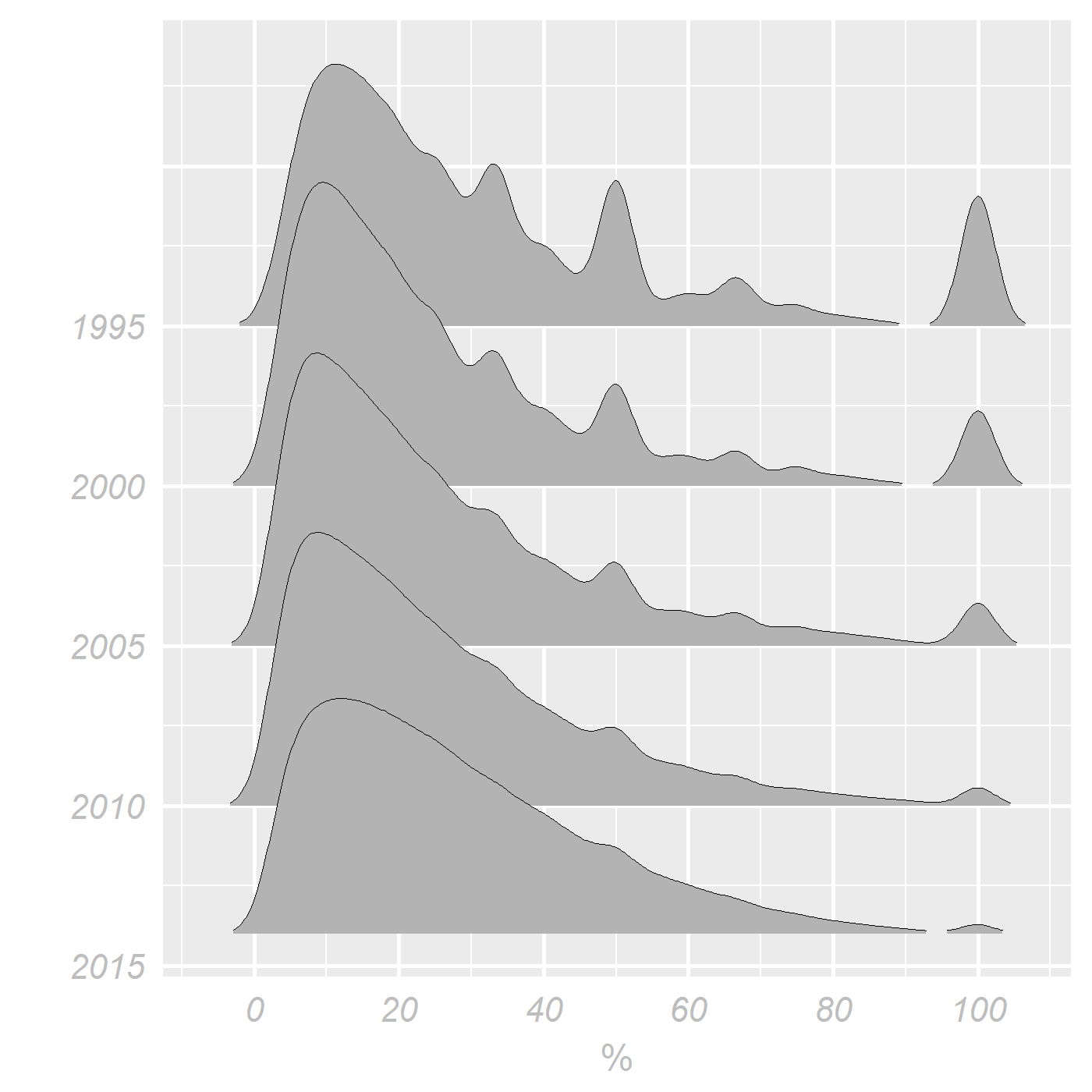} 

}

\caption[Share of Chinese national self-citations]{Share of Chinese national self-citations}\label{fig:CHN_self_cit_graph}
\end{figure}

\end{knitrout}

Figure \ref{fig:CHN_self_cit_graph} depicts the distribution of shares of references to Chinese publications by Chinese publications, i.e. the distribution of national self-citations on the level of publications. Interestingly the distribution includes several peaks at 100$\%$, 66$\%$, 50$\%$, and 33$\%$, which are vanishing as times goes by and cause the upper tail of the distribution to shrink. The lower tail of the distribution does not feature any level shift, but stays constant. Comparing the distribution of 1995 and 2014 the lower 10$\%$, 25$\%$ and 50$\%$ quantile remain at the same low level of 7$\%$, 13$\%$ and 24$\%$ of national self citations, while the 75$\%$ and 90$\%$ quantile diminish substantially by up to 20 percentage points to 40$\%$, respectively 55$\%$. Consequently other nations receive a large share of Chinese citations driving their count of obtained citations.

\begin{knitrout}
\definecolor{shadecolor}{rgb}{0.969, 0.969, 0.969}\color{fgcolor}\begin{figure}[t!]

{\centering \includegraphics[width=4in,height=3.5in,center]{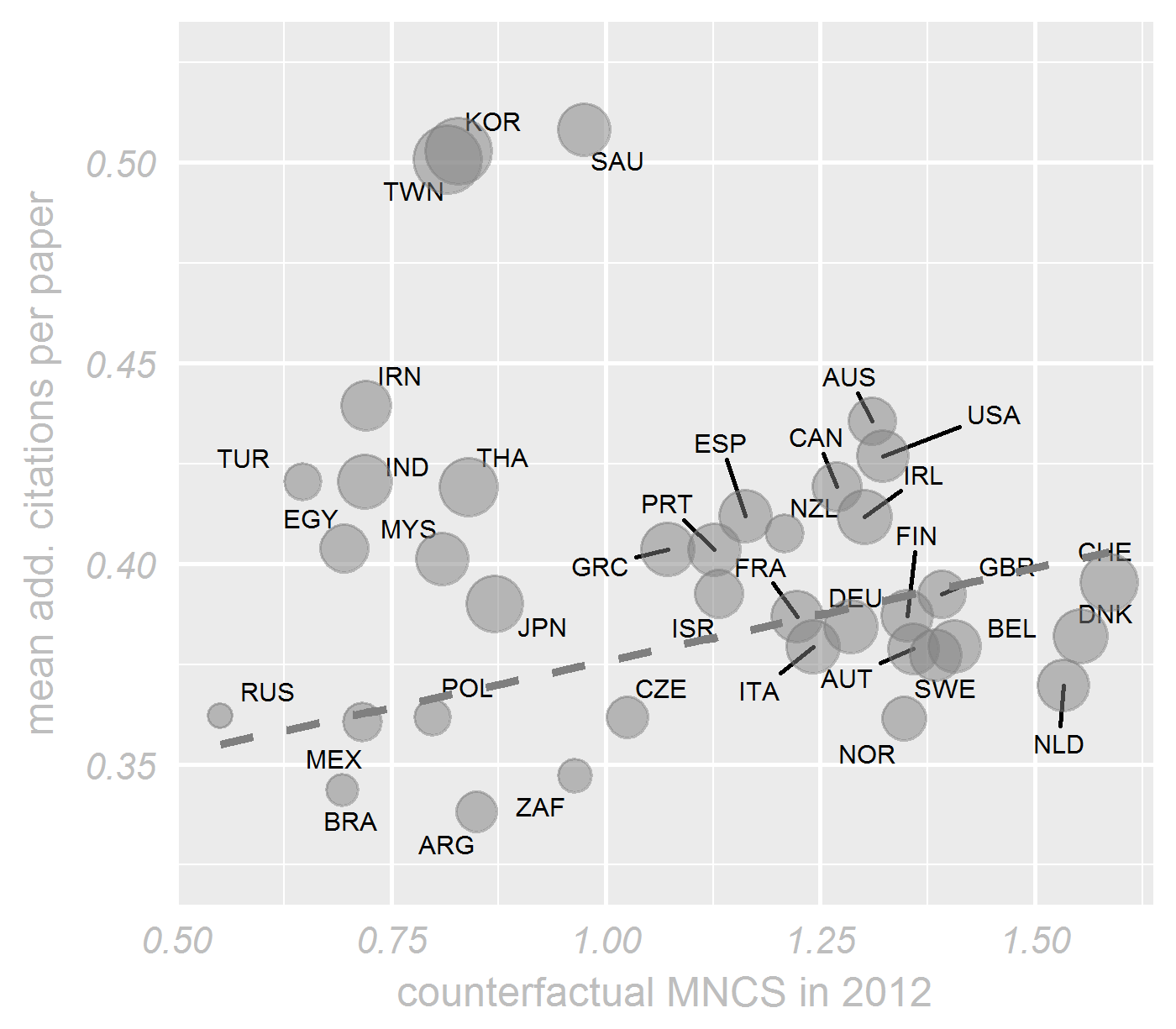} 

}

\caption[National difference between counterfactual national impact and average, normalized, additional citations from China in 2012 with correlation line for non-Asian countries]{National difference between counterfactual national impact and average, normalized, additional citations from China in 2012 with correlation line for non-Asian countries. Countries receiving  more than 1500 citations are abbreviated by three-letter codes. Singapore is omitted.}\label{fig:AdditionalCitations2012Graph}
\end{figure}

\end{knitrout}

Next we analyse how those outward citations from Chinese publications are distributed among countries. Figure \ref{fig:AdditionalCitations2012Graph} graphically relates the additional Chinese citations received by other countries in 2012 to their general scientific impact expressed via the counterfactual MNCS and their publication output. In detail the y-axis describes how many additional normalized citations a paper receives from Chinese publications on average once it is cited by a Chinese publication and the point size denotes the share of national publications being cited by Chinese publications. Jointly these measures describe the scale and intensity of Chinese citations received by other countries.

The plot groups the countries in four clusters in which we restrict our analysis to countries with at least 1500 incoming citations from Chinese publications to focus on the main effects caused by the additional Chinese citations. To the right of the Czech Republic a group of OECD countries with a MNCS larger than 1 are affected more or less alike from the Chinese citations. Nearly 30$\%$ of their articles are cited by Chinese publications and these articles obtain on average 0.4 additional normalized citations per cited publication. Countries with a MNCS below 1 can be split up in three clusters. Countries with additional citations of less than 0.375 also find around 25$\%$ of their publications being cited by Chinese publications. This group of countries, which does not include any Asian country, still benefits from the additional citations but does so less on scale and on intensity.

On the contrary the small group of countries with a MNCS of less than one, but additional citations of around 0.5 observe also a large share of nearly half of their publications being cited by Chinese publications. This small set of mainly Asian countries, namely South Korea, Taiwan, Singapore (not shown as an extreme outlier) and Saudi Arabia benefit most from Chinese citations and they do so in scale as in intensity. Several other Asian countries can be found in the last cluster which observes comparable additional citations like the established OECD countries, but exhibit a MNCS of less than 1. Consequently if we exclude the Asian countries for a moment we observe a positive correlation between MNCS and additional citations which is driven by the established OECD countries and the geographic diverse cluster of Russia, Mexico, Brasil, Poland, Argentian and South Afica. Hence Chinese publication focus their citations especially on the literature of the scientifically leading countries as these obtain higher additional citations as well as a larger share of their publications being cited. However this relation is obscured by a strong regional focus in which many Asian countries obtain more citation than stipulated by this reasoning. Yet in general developed countries seem to benefit stronger than (non-Asian) developing countries.

\begin{knitrout}
\definecolor{shadecolor}{rgb}{0.969, 0.969, 0.969}\color{fgcolor}\begin{figure}[t!]

{\centering \includegraphics[width=4.25in,height=2.5in,center]{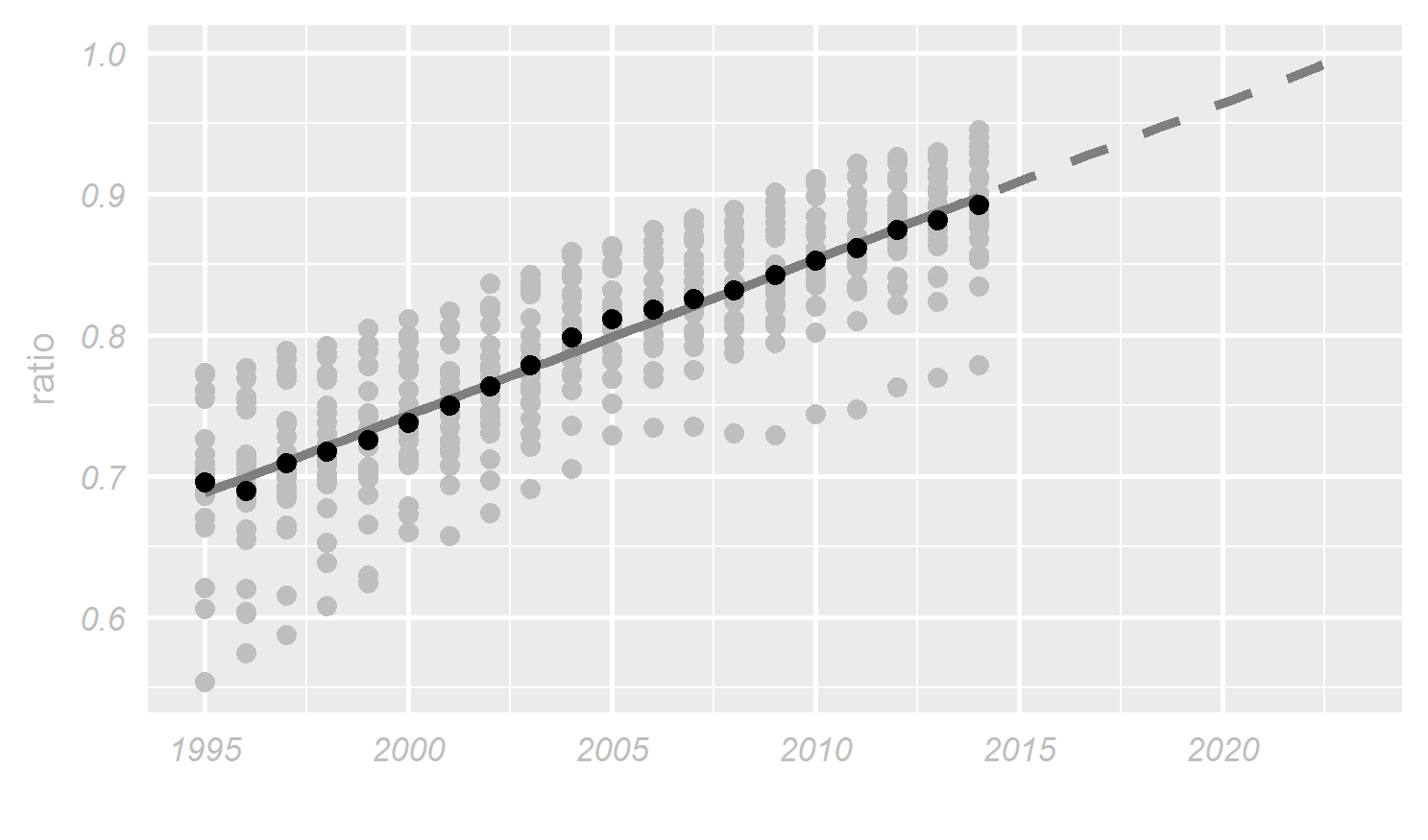} 

}

\caption[Ratio of counterfactual and factual MNCS per country (grey points), annual mean ratio (black point) and linear extrapolation]{Ratio of counterfactual and factual MNCS per country (grey points), annual mean ratio (black point) and linear extrapolation}\label{fig:CitExpCompaByCountryYearGraph}
\end{figure}

\end{knitrout}

Finally we examine how the increase in expected and obtained citations vary, as according to equation (\ref{eq2}) the MNCS and the expected citation count might only jointly rise, if the growth in obtained citations exceeds the growth in expected citations. Consequently for any country with a positive $\Delta MNCS$ the ratio of counterfactual obtained citations to actual obtained citations must be smaller than the same ratio for expected citation counts, as the increase in additional citations from Chinese publications must surpass the general increase in expected citation counts.

Such a ratio of ratios may also be derived from comparing the normalized citations $NC_i = OC_i/EC_i$ in both settings for some non--Chinese publication $i$:
\begin{alignat}{2}
~&\textstyle\frac{NC_{counterfactual}}{NC_{actual}}& ~< 1\nonumber\\
\textstyle\Leftrightarrow~&\textstyle\frac{OC_{contrafactual}}{EC_{counterfactual}}\frac{EC_{actual}}{OC_{actual}}& ~< 1\nonumber\\
\textstyle\Leftrightarrow~&\frac{\frac{OC_{contrafactual}}{OC_{actual}}}{\frac{EC_{counterfactual}}{EC_{actual}}}& ~< 1 \label{eq3}
\end{alignat}
However a direct comparison of $\Delta MNCS$ and equation (\ref{eq3}) is complicated by the fact, that the former compares the actual with the counterfactual observation in the end at a macro level, while the later conducts this comparison already in the beginning at the micro level of single publications.

While equation (\ref{eq3}) might be larger than one for some publications, it needs to be smaller for an essential share of the national publications if the corresponding $\Delta MNCS$ is to be positive. Figure \ref{fig:CitExpCompaByCountryYearGraph} depicts the country average of this publication level ratio of ratios for the aforementioned set of developed countries as grey points, while the black point depicts the average over countries. It might be noted that the ratio is on average indeed smaller than one for all aforementioned developed countries in the observation period. Consequently the number of additional citations received by these countries exceeds in general the rise in expected citation counts and their observed positive $\Delta MNCS$ rests upon this mechanism.

\section*{Outlook and implications}

Having observed the underlying mechanism, we may also ask, what will happen if the primary conditions change. The developed countries will observe a negative $\Delta MNCS$ if the formerly mentioned factors currently driving the increase in MNCS flip over. Any country not receiving citations from Chinese publications will decrease in their measured performance due to the increase in the expected counts. Furthermore a rising share of national self citations, currently unobserved in Figure \ref{fig:CHN_self_cit_graph}, or an improved reception of Chinese publication by third countries, as indirectly, but increasingly observed in the lower panel of Figure \ref{fig:CHN_bib_Graph}, will leave too few additional citations for other countries to compensate for the increase in expected citations. Consequently their MNCS will decrease although their offer on publications might stay constant and is unfavourably valued due to the changing environment of these publications.

In this respect Figure \ref{fig:CitExpCompaByCountryYearGraph} not only shows the more pronounced difference in observed citations than in expected citations between the actual and counterfactual setting, but also highlights how this gap reduces over time letting the ratio converge to 1. A simple linear time series model of the average values seem to describe a not unreasonable fit and might be employed to tentatively extrapolate from the observation period. Other things being equal this admittedly crude outlook predicts the turning point of the currently still positive effect of Chinese publications on developed countries' MNCS to occur in the not so far future.

This reasoning is also maintained by looking at the effect Chinese publications carry on the PP(top10) indicator. While the left panel in Figure \ref{fig:HCGraph} illustrates the share of highly cited publications of the aforementioned countries and does not deviate strongly from its MNCS-based equivalent, the right panel describes, likewise to the upper right panel in Figure \ref{fig:MNSCGraph}, the additional share received by these countries via the inclusion of Chinese publications. Although we will not describe the underlying mechanism, relying not only on absolute citations, but also the skewness of their distribution, in detail due to space constrains, it might be observed that the average effect curve is of parabolic shape with a maximum around 2005. Consequently the mean effect of Chinese publications on the PP(top10) values of developed countries is already declining and the confidence interval of the average effect in 2014 touches the zero line of no effect. In this regard we would expect developed countries to observe in the future at first a decline in PP(top10) shares followed by a subsequent drop in MNCS values.

\begin{knitrout}
\definecolor{shadecolor}{rgb}{0.969, 0.969, 0.969}\color{fgcolor}\begin{figure}[t]

{\centering \includegraphics[width=6.5in,height=2.5in,center]{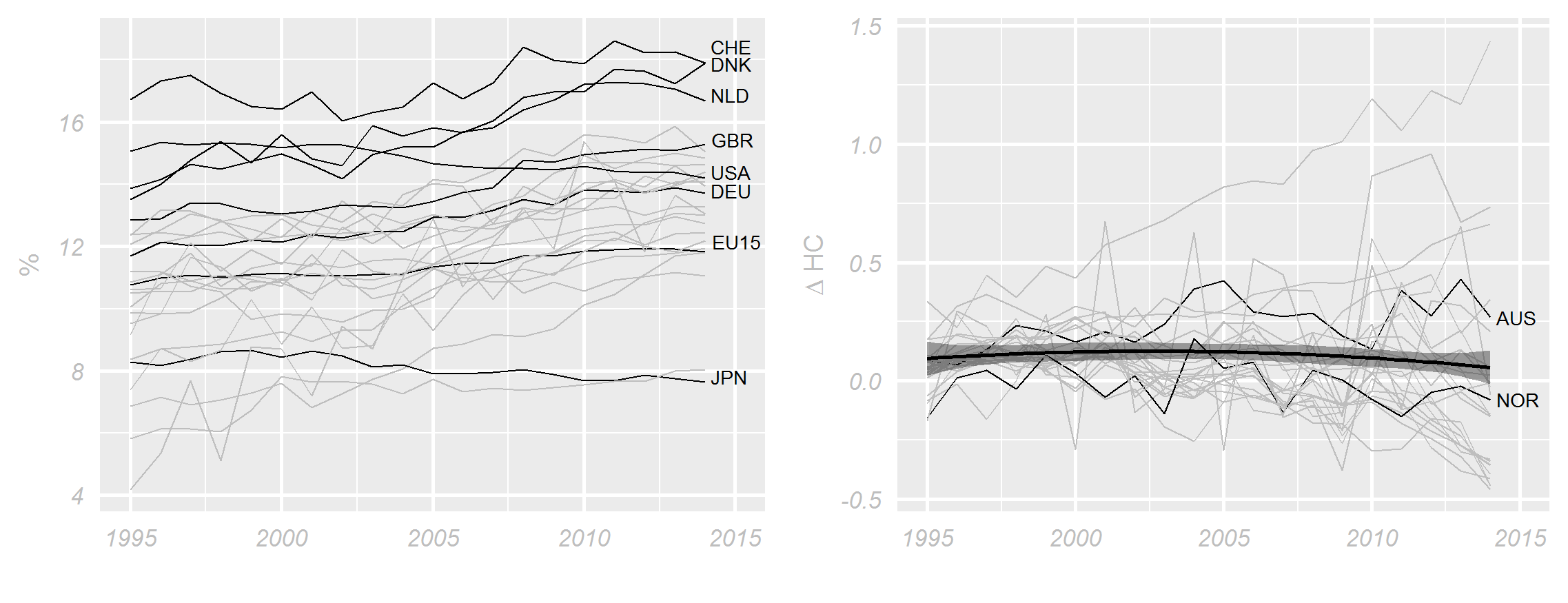} 

}

\caption[National shares of highly cited publications (left panel) and according changes due to the increase in Chinese publications (right panel)]{National shares of highly cited publications (left panel) and according changes due to the increase in Chinese publications (right panel)}\label{fig:HCGraph}
\end{figure}

\end{knitrout}

However, up to which point we might actually observe this uniform decline among developed countries is not only driven by the increasing variance of $\Delta MNCS$ and $\Delta PP(top10)$, but also by other interfering factors. The evolution of the \textit{Web of Science} publication universe is affected by several internal and external influences of which the Chinese publication increase only denotes one, albeit the probably most identifiable, effect. Other substantive changes like the intensified international collaborations resulting in multi-author publications \citep{Glaenzel2001}, the inclusion of new journals into the database \citep{Schneider2017} or the more recent open science movement \citep{Nielsen2012} all influence the setting in which a set of national publications is bibliometrically evaluated.

Consequently national bibliometric impact measures like the MNCS and the PP(top10) indicators not only mirror the national performance on its own, but also profound changes in the global environment. While every publications on its own might be understood as a change in the environment, most of these interfering influences are too small to actually alter the curse of the database. Some of these influences on the environment, however, are pronounced enough to meddle with the reference points of bibliometric evaluations and influence impact indicators. Such changes might be understood as structural and not every country will benefit in its measured bibliometric impact from them.

The tracking of national impact indicators on a macro scale is also applied to detect changes in the impact caused by varying national funding schemas and has recently gained considerable attention \citep{Schneider2016} and discussion \citep{Besselaar2017}. In light of the observed results of the Chinese publication increase and the enlisted set of further potentially structural changes in the database affecting countries to varying degrees, we like to note that the attribution of changes in the corresponding time series to national funding modifications seems to denote an especially daunting task. The upper right panel in Figure \ref{fig:MNSCGraph}, as well the right panel of Figure \ref{fig:HCGraph} show the $\Delta MNCS$, respectively the $\Delta HC$ for the in the literature analysed countries Australia and Norway. Both are clearly, if differently affected by the Chinese publication increase and it is far from obvious how an analyst might distinguish the effect caused by Chinese publications from a potentially concurrent effect driven by a modified funding schema by analysing the time series alone. Resorting to a relative comparison of time series of several countries might also not help, as countries are not uniformly affected by the Chinese publication increase. Consequently most deliberately causal analyses apply comparisons between affected and unaffected entitites in order to gauge the effect \citep{Butler2003}. However, the Chinese publication increase does not affect all entities alike and consequently constitutes a confounding mechanism for any such comparison. Controlling this nuisance denotes a potential remedy, but it's implementation in a research design might prove especially challenging, if feasible at all.

\section*{Conclusions}

We have shown how the growth of Chinese publications alters the database and carries a non-marginal effect on other nations. Publications of these countries observe changes in the respective count of obtained citations and are confronted by higher expected citations counts due to the particular bibliometric characteristics of China, namely longer reference lists and a strong focus on Asian and scientifically leading countries. Consequently national impact indicators of these countries benefit from the increasing Chinese publication output with its specific characteristics, although the upcoming reversion of these fundamental bibliometric characteristics might penalize these countries in the future.

The existence of this currently positive effect on many countries, its described origin, quantified consequences for bibliometric evaluation and list of further likewise interferences demonstrate that the publication universe finds itself in a constant state of flux. Consequently any relative measure of national impact resting upon this unsteady base informs on changes on both sides, the national as well as the global performance.

This ascertainment entails important consequences for the interpretation of national impact values, as the direct link between the national performance and the impact value is partially impaired by structural changes in the environment. Controlling these confounding mechanisms seems essential to an unbiased measurement of national bibliometric performance.

\end{document}